\def\year{2020}\relax
\documentclass[letterpaper]{article} 
\usepackage{aaai20}  
\usepackage{times}  
\usepackage{helvet} 
\usepackage{courier}  
\usepackage[hyphens]{url}  
\usepackage{graphicx} 
\usepackage{graphicx}  
\usepackage{amsmath,amssymb,tikz,pgfplots,verbatim,xcolor,xspace,multirow,booktabs,subfigure}
\usepackage{pifont}
\usepackage{stix}
\usepackage[multiple]{footmisc}
\newcommand{\cmark}{\ding{51}}%
\newcommand{\xmark}{\ding{53}}%
\pgfplotsset{compat=1.3}

\title{Deep Residual-Dense Lattice Network for Speech Enhancement}

\author{
\Large \textbf{Mohammad Nikzad,\textsuperscript{\rm 1} Aaron Nicolson,\textsuperscript{\rm 1} Yongsheng Gao,\textsuperscript{\rm 1}}\\
\Large \textbf{Jun Zhou,\textsuperscript{\rm 1} Kuldip K. Paliwal,\textsuperscript{\rm 1} Fanhua Shang\textsuperscript{\rm 2}}\\
\textsuperscript{\rm 1}Institute for Integrated and Intelligent Systems, Griffith University, Australia\\ 
\textsuperscript{\rm 2} School of Artificial Intelligence, Xidian University, China\\
$\{$m.nikzadehaji,  aaron.nicolson$\}$@griffithuni.edu.au, \\
$\{$yongsheng.gao, jun.zhou, k.paliwal$\}$@griffith.edu.au\\
 fhshang@xidian.edu.cn
}

\begin{document}

\maketitle

\begin{abstract}
Convolutional neural networks (CNNs) with residual links (ResNets) and causal dilated convolutional units have been the network of choice for deep learning approaches to speech enhancement. While residual links improve gradient flow during training, feature diminution of shallow layer outputs can occur due to repetitive summations with deeper layer outputs. One strategy to improve feature re-usage is to fuse both ResNets and densely connected CNNs (DenseNets). DenseNets, however, over-allocate parameters for feature re-usage. Motivated by this, we propose the \textit{residual-dense lattice} network (RDL-Net), which is a new CNN for speech enhancement that employs both residual and dense aggregations without over-allocating parameters for feature re-usage. This is managed through the topology of the RDL blocks, which limit the number of outputs used for dense aggregations. Our extensive experimental investigation shows that RDL-Nets are able to achieve a higher speech enhancement performance than CNNs that employ residual and/or dense aggregations. RDL-Nets also use substantially fewer parameters and have a lower
computational requirement. Furthermore, we demonstrate that RDL-Nets outperform many state-of-the-art deep learning approaches to speech enhancement.\\
\textit{Availability:} https://github.com/nick-nikzad/RDL-SE.

\end{abstract}

\section{Introduction}

Deep learning approaches to speech enhancement represent a significant leap in performance over previous approaches, such as the decision-directed (DD) approach \cite{1164453}. Multi-layer perceptrons (MLPs) were amongst the first artificial neural networks (ANNs) used for speech enhancement \cite{xu2017multi}. Recurrent neural networks (RNNs) employing long short-term memory (LSTM) cells provided a higher performance at the cost of parameter inefficiency and extensive training times \cite{chen2017long}. Convolutional neural networks (CNNs) were able to match the performance of LSTM networks, with fewer parameters and a reduction in training time \cite{Park2017}. LSTM networks were not outperformed until the introduction of residual \cite{he2016identity,wavenet} and densely connected \cite{huang_densely_2017,dense_SE} CNNs, as well as causal dilated convolutional units \cite{bai2018empirical}. A residual CNN (ResNet) aggregates layer outputs via a summation operation, which is given as input to deeper layers. A densely connected CNN (DenseNet) differs by aggregating layer outputs via a concatenation operation. Other ANNs that have been successfully applied to speech enhancement include generative adversarial networks (GANs) and encoder-decoder CNNs \cite{segan}.

\begin{figure}[htp]
\centering
	\centering
	\includegraphics[scale=1.1]{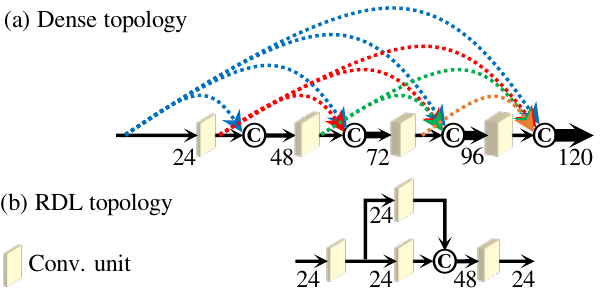}%
	\caption{Comparison of (a) the dense topology and (b) the proposed RDL topology. The number of input features to each convolutional unit is indicated. Given identical kernel and output sizes, more parameters are consumed for a larger input size. \textcircled{c} represents the concatenation operation.}
	\label{fig:res_dense_block}
\end{figure}

Residual and dense aggregations of layer outputs have been found to benefit training. Residual links improve gradient flow during backpropagation \cite{he2016identity} and prevent the vanishing and exploding gradient problems \cite{279181}. This allows the training of very deep neural networks. Dense aggregations offer direct feature re-usage, as deeper layers have access to the outputs of shallower layers \cite{huang_densely_2017}. This allows a layer to explore a larger set of features during training. Despite the success of both ResNets and DenseNets, both aggregation types have drawbacks. For ResNets, information from the outputs of shallower layers can be lost after multiple summations with deeper layer outputs \cite{sparse_aggCNN}. This restricts feature re-usage and limits feature exploration during training. For DenseNets, while the concatenation of all previous layer outputs yields total feature re-usage, a significant number of parameters are unexploited due to large input sizes \cite{MLN}. This is exemplified in Figure \ref{fig:res_dense_block} (a), where the input size increases with the depth of the block.

\begin{figure}[t]
\centering
	\centering
	\includegraphics[scale=0.67]{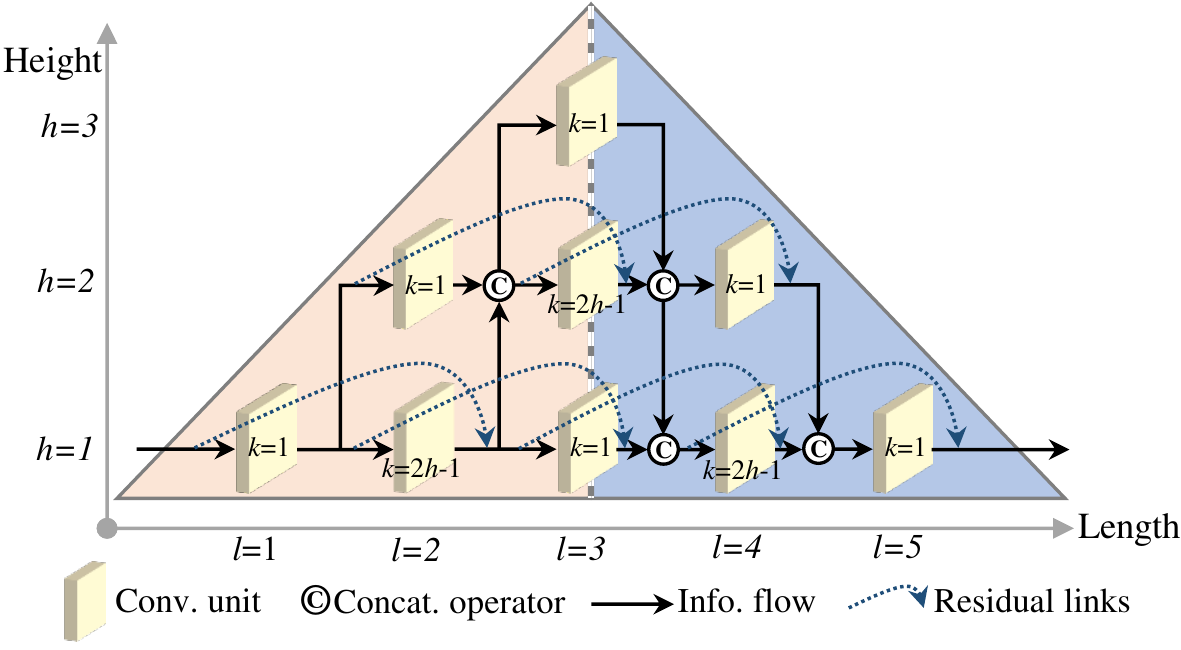}%
	\caption{An RDL block with a length of 5 and height of 3. The two coloured triangles indicate the left and right halves of the lattice. Here, the kernel size is denoted by $k$.}
	\label{fig:model_block}
\end{figure}

Combining the benefits of both aggregation types has also been investigated. Mixed link networks (MLNs) are CNNs that employ both residual and dense aggregations \cite{MLN}. A network that employs densely connected residual blocks (DenseRNet) was able to outperform both ResNets and DenseNets on a speech recognition task \cite{DenseRNet}. As shown in \cite{MLN}, MLNs such as DenseRNets follow a similar dense aggregation strategy to DenseNets. For example, DenseRNet blocks have total feature re-usage between residual blocks. This indicates that current MLNs possess the same drawback inherent with DenseNets: too many parameters are allocated for feature re-usage in each block.
In this paper, we propose a new CNN for speech enhancement that takes advantage of both aggregation types, without over-allocating parameters for feature re-usage. This is achieved by using a topology that differs from the chain-structure of MLNs, such as DenseRNet. The topology is a triangular lattice of convolutional units, as illustrated in Figure \ref{fig:model_block}. Local dense aggregations of convolutional unit outputs are formed strictly over the \textit{height} of the lattice. As can be seen by comparing Figure 1 (b) to Figure 1 (a), this reduces the maximum input size to a convolutional unit within a block. While RDL blocks do not allow for total feature re-usage, densely aggregating only a subset of previous outputs has been shown to be beneficial \cite{sparse_aggCNN}. Local residual and global dense links are also adopted, to improve intra block gradient flow, and inter block feature re-usage, respectively. We refer to the framework of applying residual and dense aggregations over a triangular lattice of convolutional units as a {\it residual-dense lattice} (RDL). Moreover, we show that the proposed RDL network (RDL-Net) is able to produce a higher speech enhancement performance than networks that employ residual and/or dense aggregations. An ablation study of RDL-Nets is also performed over multiple aggregation configurations. We also show that RDL-Nets outperform many state-of-the-art deep learning approaches to speech enhancement.

\section{Related works}\label{sec_rlw}
ANNs have been used for enhancing speech in both the time- and frequency-domain. In the time-domain, ANNs estimate clean speech frames from given noisy speech frames. A GAN was employed for speech enhancement in the time-domain (SEGAN), which used encoder-decoder CNNs for both the generator and discriminator \cite{segan}. A CNN employing non-causal dilated convolutional units and residual links was also used for speech enhancement in the time-domain (Wavenet) \cite{wavenet}.

In the frequency-domain, ANNs are employed to estimate either the clean speech magnitude spectra, a time-frequency mask, or the \textit{a priori} SNR from given noisy speech magnitude spectra. An MLP was used to estimate the clean speech log-power spectra (LPS) \cite{6932438}, with the framework later incorporating multi-objective learning, and ideal binary mask (IBM) post-processing (Xu2017) \cite{xu2017multi}. A DenseNet was also used to estimate the clean speech LPS in \cite{dense_SE}, and was able to outperform both MLP and LSTM networks in the same framework. Time-frequency masks, such as the ideal ratio mask (IRM), are applied as a suppression function to the noisy speech magnitude spectra. An LSTM network was used to estimate the IRM (LSTM-IRM) \cite{chen2017long}, which was able to generalise to unseen speakers. A GAN with a regularised loss function was also used to estimate the IRM (MMSE-GAN), and was able to outperform SEGAN \cite{MMSE-GAN}. This was outperformed by another GAN IRM estimator that used multiple objective measures during optimisation (Metric-GAN) \cite{metric-GAN}.

\textit{A priori} SNR estimates are used by minimum mean-square error (MMSE) estimators of the clean speech magnitude spectra \cite{1164453}. Recently, a deep learning approach to \textit{a priori} SNR estimation was proposed (Deep Xi) \cite{nicolson2019deep}. It used a residual LSTM network (ResLSTM) to estimate the \textit{a priori} SNR directly from noisy speech magnitude spectra. By estimating the \textit{a priori} SNR, different MMSE approaches can be used such as the MMSE log-spectral amplitude (MMSE-LSA) estimator \cite{1164550} and the square-root Wiener filter (SRWF) \cite{1455809}. RDL-Nets are examined within the Deep Xi framework, due to its flexibility of MMSE estimator choice.
\begin{figure*}[ht]
	\centering 
	\includegraphics[scale=0.83]{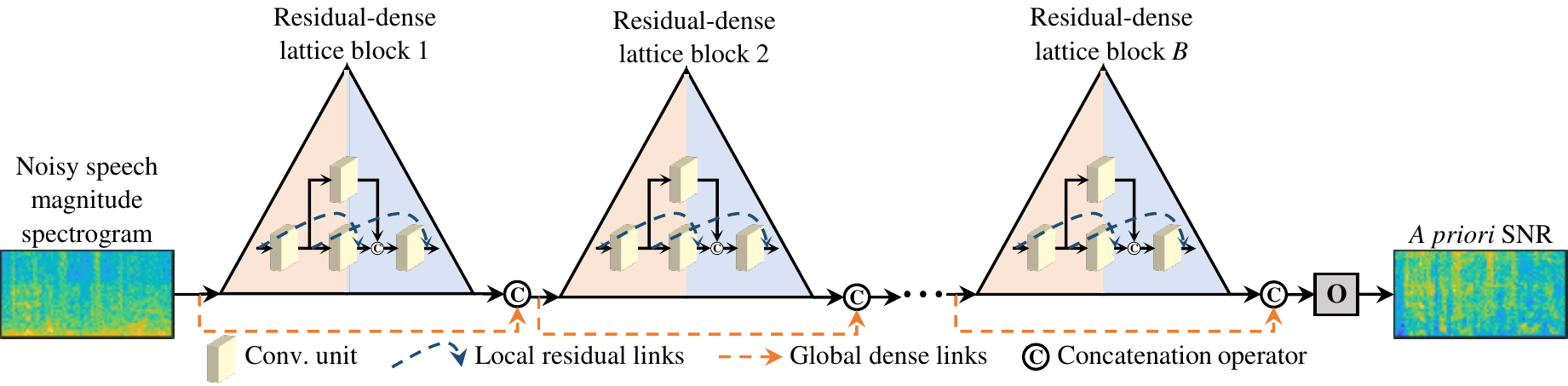}
	\caption{The proposed RDL-Net for speech enhancement. The RDL-Net estimates the \textit{a priori} SNR from the given noisy speech magnitude spectrum. The estimated \textit{a priori} SNR is then used by an MMSE clean speech magnitude spectrum estimator.}
	\label{fig:model_network}
\end{figure*}
\section{Residual-dense lattice networks} \label{seca}

The proposed RDL-Net is used to estimate the \textit{a priori} SNR from given noisy speech magnitude spectra, as shown in Figure \ref{fig:model_network}. The network consists of $B$ RDL blocks, and a sigmoidal fully-connected output layer, $\textbf{O}$. The block topology is a triangular lattice of convolutional units, as shown in Figure \ref{fig:model_block}. The location of each convolutional unit within the lattice, $C_{hl}$, is specified by a height and length co-ordinate $(h,l)$, where $h=1,2,...,H$, and, $l=1,2,...,L$. The number of convolutional units in each block is denoted by $N$, where $N$ is a square number, and $N\geq 4$. The height of the lattice is $H=\sqrt{N}$, and the length is $L=2\sqrt{N}-1$. The following notation is used to indicate in which section of the lattice a convolutional unit exists:

\begin{equation}
C_{hl}= \begin{cases}
C^{\lrtriangle}_{hl},& \text{if }h\leq l,~ l\leq H \\
C^{\lltriangle}_{hl},& \text{if }h \leq 2H-l,~ H<l\leq L \\
\emptyset,& \text{otherwise},\\
\end{cases}
\end{equation}
where $C^{\lrtriangle}_{hl}$ and $C^{\lltriangle}_{hl}$ are convolutional units that exist in the left and right triangles of the lattice, respectively. $\emptyset$ indicates that the convolutional unit does not exist. 

\subsection{Convolutional units}
Each convolutional unit is a composite function, $f(\cdot)$, consisting of three operations, including layer normalisation \cite{ba2016layer}, followed by ReLU activation \cite{glorot_deep_2011}, and 1D causal dilated convolution \cite{bai2018empirical}. The output of a convolutional unit is given by $f(x_{hl},W_{hl})$, where $W_{hl}$ denotes the weights (and biases). Convolutional units within the RDL-Net are connected by both local and global links (i.e. intra and inter block links).

\subsection{Local dense aggregations}
The input to a convolutional unit in the left triangle of the lattice, $x^{\lrtriangle}_{hl}$, is the dense aggregation of the outputs at length $l-1$, and heights $h,h-1,...,1$:
\begin{equation}
\begin{aligned}
x^{\lrtriangle}_{hl} &= \begin{cases}
x_{11}, & \text{if } l=h=1\\
y_{h(l-1)}, & \text{if  } C^{\lrtriangle}_{hl} ,~ l>1, h=1\\
[y_{h(l-1)},x_{(h-1)l}], & \text{if  } C^{\lrtriangle}_{hl} ,~ l>h, h>1\\
x_{(h-1)l}, & \text{if  } C^{\lrtriangle}_{hl} ,~ l=h, h>1,
\end{cases}
\end{aligned}
\end{equation}
where $[.]$ denotes the concatenation operation, and $y_{hl}$ is the local residual aggregation at $(h,l)$. The local dense aggregations in the left triangle of the lattice allow for multiple concise outputs to be progressively formed. The input to a convolutional unit in the right triangle of the lattice, $x^{\lltriangle}_{hl}$, is the dense aggregation of the outputs at length $l-1$, and heights $h,h+1,...,H$:
\begin{equation}
\begin{aligned}
x^{\lltriangle}_{hl} &= \begin{cases}
[y_{h(l-1)},y_{(h+1)(l-1)}], & \text{if } C^{\lltriangle}_{hl} ,~ h=2H-l\\
[y_{h(l-1)},x_{(h+1)l}], & \text{if } C^{\lltriangle}_{hl} ,~ h<2H-l.
\end{cases}
\end{aligned}
\end{equation}
In the right triangle of the lattice, the outputs are progressively amalgamated into a single output. By densely aggregating outputs over the height of the lattice, the input size to deeper convolutional units within the block is limited. This enables RDL-Nets to avoid the drawback associated with other densely connected residual networks: \textit{the over-allocation of parameters for feature re-usage}.

\subsection{Local residual aggregations}
To improve the flow of gradients over the length of the lattice, local residual links are adopted:
\begin{equation}
y_{hl}= \begin{cases}
f(x_{hl},W_{hl})+x_{h(l-1)}, & \text{if  } C^{\lrtriangle}_{hl}\text{ or }C^{\lltriangle}_{hl},~l>h\\
f(x_{hl},W_{hl}), & \text{if  } C^{\lrtriangle}_{hl}\text{ or }C^{\lltriangle}_{hl},~ l\leq h.\\
\end{cases}
\end{equation}
When the size of $y_{hl}$ and $x_{h(l-1)}$ are non-identical, the residual link is weighted so that $x_{h(l-1)}$ is the same size as $y_{hl}$. Local residual links also help to stabilise the training process~\cite{he2016identity}.

\subsection{Global dense aggregations}
Global dense links are adopted, to further enhance the propagation of information between RDL blocks:
\begin{equation}
x^{b+1}_{11}=[x^{b}_{11}, y^{b}_{1L}],
\end{equation}
where the superscript is added to the notation to indicate the block index, $b=1,2,...,B$. Utilising global dense links also enables feature re-usage between the RDL blocks. 

\subsection{Implementation details}
The receptive field of an RDL block is controlled via the dilatation rate, $d=2^{h-1}$, and the kernel size, $k=2h-1$. However, this strategy can expend a large number of parameters. Hence, we alternate the kernel size of $k=2h-1$, with $k=1$ at each length, as depicted in Figure \ref{fig:model_block}. Moreover, we set the convolutional unit output size at each height to $m_{h}=\frac{m_1}{2^{h-1}}$, where $m_1$ is the output size at $h=1$. This policy ensures that a reduced number of parameters are used for feature re-usage. In this work, the total number of convolutional units for each RDL block was set to $N=16$ (hence, $H=4$ and $L=7$). The output size of the first level ($h=1$) was $m_{1}=64$. RDL-Nets with sizes of 0.53, 1.08, 1.48, 1.87, and 3.91 million parameters were formed by cascading 3, 6, 8, 10, and 18 blocks, respectively.

\section{Experiment setup} \label{secb}
\subsection{Network Configurations}
The aforementioned RDL-Net configurations and the following network configurations were tasked with estimating the \textit{a priori} SNR within the Deep Xi framework \cite{nicolson2019deep}. The estimated \textit{a priori} SNR is then used by MMSE approaches to speech enhancement.
\begin{description}
	\item[ResNet:] Each residual block contained 2 causal dilated convolutional units with an output size of 64, and a kernel size of 3. For each block, $d$ was cycled from 1 to 8 (increasing by a power of 2). ResNets of sizes 0.53, 1.03, 1.53, and 2.03 million parameters were formed by cascading 20, 40, 60, and 80 residual blocks, respectively.
	
	\item[DenseNet:] Each dense block contained 4 causal dilated convolutional units with an output size of 24, and a kernel size of 3. For each convolutional unit, $d$ was cycled from 1 to 8 (increasing by a power of 2). DenseNets of sizes 0.57, 0.97, 1.48, and 2.10 million parameters were formed by cascading 5, 7, 9, and 11 dense blocks, respectively.
	
	\item[DenseRNet:] Each denseR block was composed of 4 densely connected residual blocks. Each residual block contained 2 causal dilated convolutional units with an output size of 24 and a kernel size of 3. For each residual block, $d$ was cycled from 1 to 8 (increasing by a power of 2). DenseRNets of sizes 0.60, 1.05, 1.44, and 2.02 million parameters were formed by cascading 2, 3, 4, and 6 denseR blocks, respectively.	
	
	\item[ResLSTM:] The cell size and number of residual blocks were 170 and 4, 188 and 5, and 200 and 6, for the ResLSTMs of sizes 1.02, 1.51, and 2.03 million parameters, respectively. This was the original network used in the Deep Xi framework \cite{nicolson2019deep}.
\end{description}

\subsection{Speech enhancement}
For each frame of noisy speech, the 257-point single-sided magnitude spectrum was computed, which included both the DC frequency component and the Nyquist frequency component, forming the input to each of the five previously described networks. The estimated \textit{a priori} SNR was used by an MMSE approach (MMSE-LSA estimator or SRWF approach) to estimate the clean speech magnitude spectrum. The short-time Fourier analysis, modification, and synthesis (AMS) framework was used to produce the final enhanced speech \cite{nicolson2019deep}. The Hamming window function was used for analysis and synthesis, with a frame length of 32 ms and a frame shift of 16 ms. 

\subsection{Training set}
The \textit{train-clean-100} set from the Librispeech corpus \cite{panayotov2015librispeech}, the CSTR VCTK corpus (recordings from speakers $p232$ and $p257$ were excluded as they are used in Test Set 2) \cite{veaux2017cstr}, and the $si^*$ and $sx^*$ training sets from the TIMIT corpus \cite{garofolo1993darpa} were included in the training set ($73\,404$ clean speech recordings). $5\%$ of the clean speech recordings ($3\,667$) were randomly selected and used as the validation set. The $2\,382$ recordings adopted in \cite{nicolson2019deep} were used for the noise training set. All clean speech and noise recordings were single-channel, with a sampling frequency of 16 kHz. The noise corruption procedure for the training set is described in the next subsection.

\subsection{Training strategy}\label{sece}
Cross-entropy was used as the loss function. The \textit{Adam} algorithm \cite{kingma2014adam} with default hyper-parameters was used for stochastic gradient descent optimisation. A mini-batch size of $10$ noisy speech signals was used. The noisy speech signals were generated as follows: each clean speech recording selected for a mini-batch was mixed with a random section of a randomly selected noise recording at a randomly selected SNR level (-10 to 20 dB, in 1 dB increments). A total of 100 epochs were use to train all CNN architectures. A total of 10 epochs were used for the ResLSTM networks and the LSTM-IRM estimator \cite{chen2017long}, as each epoch required eight hours of training.
\subsection{Test sets}
The following two datasets were used for testing:
\begin{itemize}
\item {\textbf{Test set 1}:} The first test set was used to obtain the results in Figures~\ref{fig:abl-Curves} and~\ref{fig:Curves}, and Tables~\ref{table_ablstudy},~\ref{taba}, and~\ref{tabb}. The four noise sources included \textit{voice babble}, \textit{F16}, and \textit{factory} from the RSG-10 noise dataset \cite{steeneken1988description} and \textit{street music} (recording no. $26\,270$) from the Urban Sound dataset \cite{salamon2014dataset}. 10 clean speech recordings were randomly selected (without replacement) from the TSP speech corpus \cite{kabal2002tsp} for each of the four noise recordings. To generate the noisy speech, a random section of the noise recording was mixed with the clean speech at the following SNR levels: -5 to 15 dB, in 5 dB increments. This created a test set of 200 noisy speech signals. The noisy speech was single channel, with a sampling frequency of 16 kHz.
\item {\textbf{Test set 2}:} The second test set was used to obtain the results in Table \ref{table-stoa}. In order to make a direct comparison, the second test set is identical to those used in previous works. The test set included 824 clean speech recordings of two speakers from the Voice Bank corpus (393 from $p232$ and 431 from $p257$) \cite{6709856}. A total of 20 different conditions were used to create the noisy speech, including five noise types from the DEMAND dataset \cite{thiemann2013diverse}, and four SNR levels: 2.5, 7.5, 12.5, and 17.5 dB. This corresponds to approximately 20 different
sentences per condition for each speaker (824 noisy speech signals in the second test set). 
\end{itemize}

\section{Results and discussion} \label{secc}
\subsection{Local and global aggregation study}\label{subsec:abls}
In this section we conduct an ablation study on the effects of two aggregation types used in the RDL-Net topology, including local residual links (LR) and global dense links (GD). To this end, four RDL-Net configurations are examined, as shown in Table~\ref{table_ablstudy}. The convergence of each configuration during training is also depicted in Figure~\ref{fig:abl-Curves}. The four configurations were formed using the aforementioned hyper-parameters, with 5 blocks. By adding either LR or GD to the baseline (no LR or GD), it can be seen that a lower validation error can be attained. While GD aggregations add more trainable parameters (0.23M) to the baseline, it achieved a lower validation error than LR (141.82 vs. 142.14). However, the GD configuration caused obvious fluctuations in the validation error during training. Utilising both LR and GD produced the lowest validation error, without the fluctuations in validation error exhibited by the GD configuration. This demonstrates that enhanced intra block gradient flow and inter block feature re-usage are both highly beneficial to the training of an RDL-Net.

\begin{table}[t]
\centering
\caption{Ablation study of local residual (LR) and global dense (GD) aggregations.}
\begin{tabular}{c|c|c|c|c} 
 \hline
 \multicolumn{5}{c}{Different combinations of LR, GD} \\ [0.5ex] 
 \hline\hline
 LR & \xmark & \cmark& \xmark & \cmark\\ 
 GD & \xmark & \xmark & \cmark & \cmark\\ [0.5ex]
 \hline\hline
 \# params.& 0.39M & 0.53M & 0.62M & 0.86M \\\hline
 Val. error & 146.37& 142.14& 141.82& 141.43 \\
 \hline
\end{tabular}
\label{table_ablstudy}
\end{table}

\definecolor{darkgreen}{RGB}{0,128,0}
\begin{figure}[t]
\centering
\begin{tikzpicture}[scale=0.65]
\pgfplotstableread{abl_results.txt}
    \datatable
    \begin{axis}[grid=major,ymin=141.3,ymax=149,xmin=2,xmax=100,xlabel= Epoch,legend pos=outer north east,
    ylabel= Validation error,
    height=5.5cm, width=7cm]
    \addplot[color=black, densely dotted, line width=1pt] table[y = ARF] from \datatable ;
	\addlegendentry{Baseline}

    \addplot[color=darkgreen, line width=1pt, densely dashdotted] table[y = RL-ARF] from \datatable ;
	\addlegendentry{LR}	
	
    \addplot[color=blue, line width=1pt, densely dashed] table[y = GD-ARF] from \datatable ;
	\addlegendentry{GD}

	\addplot[color=red, line width=1pt] table[y = all] from \datatable ;
	\addlegendentry{LR-GD}
	
	\end{axis}
\end{tikzpicture}
\caption{Validation error attained by four RDL-Net configuration types: Baseline, LR, GD, and LR-GD.}
	\label{fig:abl-Curves}
\end{figure}
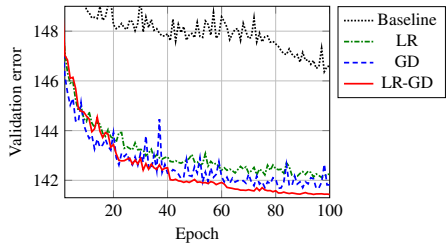

\begin{table*}[ht]
	\centering
	\scriptsize
	\setlength{\tabcolsep}{3.6pt}
	\caption{Enhanced speech objective quality scores. The mean opinion score of the listening quality objective (MOS-LQO) was used as the metric, where the wideband perceptual evaluation of quality (Wideband PESQ) was the objective model used to obtain the MOS-LQO score \cite{itut2007application}. The tested conditions include clean speech mixed with real-world \textbf{non-stationary} (\textit{voice babble} and \textit{street music}) and \textbf{coloured} (\textit{F16} and \textit{factory}) noise sources at multiple SNR levels. The highest MOS-LQO score attained at each condition and for each parameter size is shown in boldface. The standard error (SE) over all conditions for each network is provided in the last column.}
	\begin{tabular}{ll|lllll|lllll|lllll|lllll|l} 
		\toprule
		
		\multirow{3}{*}{\begin{tabular}[c]{@{}c@{}}\textbf{Network}\end{tabular}} &
		
		\multirow{3}{*}{\begin{tabular}[c]{@{}c@{}}\textbf{\# params.}\\$\pmb{\times 10^{6}}$\end{tabular}} & 
		
		\multicolumn{20}{c|}{\bf SNR level (dB)} & \multirow{3}{*}{\begin{tabular}[c]{@{}c@{}}\textbf{SE}\end{tabular}} \\ 
		\cline{3-22}
		& & \multicolumn{5}{c|}{\bf Voice babble} & \multicolumn{5}{c|}{\bf Street music} & \multicolumn{5}{c|}{\bf F16} & \multicolumn{5}{c|}{\bf Factory} &  \\ 
		\cline{3-22}
		& & {\bf-5}  & {\bf0}   & {\bf5}   & {\bf10}  & {\bf15}        & {\bf-5}  & {\bf0}   & {\bf5}   & {\bf10}  & {\bf15}        & {\bf-5}  & {\bf0}   & {\bf5}   & {\bf10}  & {\bf15}  & {\bf-5}  & {\bf0}   & {\bf5}   & {\bf10}  & {\bf15} &           \\ 
		\hline
		
Noisy speech & - & 1.04 & 1.07 & 1.15 & 1.35 & 1.71 & 1.04 & 1.06 & 1.11 & 1.27 & 1.58 & 1.03 & 1.06 & 1.11 & 1.25 & 1.52 & 1.04 & 1.04 & 1.09 & 1.24 & 1.54 & 0.016 \\
\midrule
ResLSTM & 1.02 & 1.07 & 1.20 & 1.51 & 1.96 & 2.49 & 1.10 & 1.23 & 1.50 & 1.92 & 2.47 & 1.13 & 1.28 & 1.49 & 1.80 & 2.33 & 1.09 & 1.22 & 1.54 & 1.86 & 2.31 & 0.035 \\
DenseNet & 0.97 & 1.08 & 1.20 & 1.54 & 2.01 & 2.50 & 1.09 & 1.24 & 1.51 & 1.86 & 2.36 & 1.14 & 1.35 & 1.64 & 2.05 & 2.45 & 1.07 & 1.25 & 1.55 & 1.97 & 2.43 & 0.037 \\
DenseRNet & 1.05 & 1.07 & 1.21 & 1.50 & 1.92 & 2.43 & 1.11 & 1.22 & 1.48 & 1.87 & 2.37 & 1.14 & 1.34 & 1.57 & 1.93 & 2.34 & 1.07 & 1.21 & 1.44 & 1.85 & 2.36 & 0.035 \\
ResNet & 1.03 & 1.08 & 1.22 & 1.58 & 2.10 & \textbf{2.67} & 1.11 & 1.30 & 1.58 & 2.04 & \textbf{2.53} & 1.19 & 1.44 & \textbf{1.78} & \textbf{2.17} & 2.58 & 1.11 & 1.29 & 1.60 & 2.03 & 2.46 & 0.040 \\
Prop. RDL-Net & 1.08 & \textbf{1.10} & \textbf{1.29} & \textbf{1.65} & \textbf{2.15} & 2.62 & \textbf{1.13} & \textbf{1.32} & \textbf{1.66} & \textbf{2.11} & \textbf{2.53} & \textbf{1.25} & \textbf{1.48} & 1.73 & \textbf{2.17} & \textbf{2.62} & \textbf{1.15} & \textbf{1.39} & \textbf{1.73} & \textbf{2.10} & \textbf{2.54} & 0.039 \\
\midrule
ResLSTM & 1.51 & 1.09 & 1.25 & 1.56 & 2.03 & 2.47 & 1.09 & 1.26 & 1.56 & 1.95 & 2.44 & 1.16 & 1.35 & 1.60 & 1.87 & 2.19 & 1.11 & 1.30 & 1.60 & 1.94 & 2.35 & 0.037 \\
DenseNet & 1.41 & 1.06 & 1.19 & 1.51 & 1.96 & 2.49 & 1.10 & 1.23 & 1.50 & 1.87 & 2.31 & 1.14 & 1.35 & 1.63 & 1.99 & 2.43 & 1.09 & 1.30 & 1.63 & 2.00 & 2.44 & 0.036 \\
DenseRNet & 1.37 & 1.07 & 1.22 & 1.54 & 2.00 & 2.50 & 1.12 & 1.29 & 1.59 & 1.99 & 2.45 & 1.20 & 1.41 & 1.71 & 2.10 & 2.53 & 1.06 & 1.24 & 1.57 & 2.00 & 2.45 & 0.038 \\
ResNet & 1.53 & 1.08 & 1.25 & 1.61 & 2.12 & 2.64 & 1.10 & 1.28 & 1.56 & 2.00 & 2.48 & 1.18 & 1.41 & 1.72 & 2.15 & \textbf{2.61} & 1.10 & 1.30 & 1.64 & 2.07 & 2.53 & 0.040\\
Prop. RDL-Net & 1.48 & \textbf{1.12} & \textbf{1.31} & \textbf{1.67} & \textbf{2.20} & \textbf{2.75} & \textbf{1.17} & \textbf{1.41} & \textbf{1.75} & \textbf{2.09} & \textbf{2.59} & \textbf{1.28} & \textbf{1.52} & \textbf{1.85} & \textbf{2.25} & \textbf{2.61} & \textbf{1.17} & \textbf{1.40} & \textbf{1.74} & \textbf{2.12} & \textbf{2.59} & 0.040 \\
\midrule
ResLSTM & 2.03 & 1.09 & 1.23 & 1.51 & 2.02 & 2.48 & 1.13 & 1.30 & 1.59 & 2.06 & 2.50 & 1.19 & 1.37 & 1.61 & 1.92 & 2.29 & 1.14 & 1.35 & 1.64 & 2.01 & 2.48 & 0.036 \\
DenseNet & 1.94 & 1.07 & 1.21 & 1.54 & 2.04 & 2.47 & 1.09 & 1.21 & 1.46 & 1.86 & 2.33 & 1.16 & 1.36 & 1.65 & 1.98 & 2.44 & 1.09 & 1.27 & 1.64 & 2.03 & 2.49 & 0.037 \\
DenseRNet & 2.02 & 1.08 & 1.20 & 1.48 & 1.83 & 2.24 & 1.10 & 1.21 & 1.42 & 1.77 & 2.23 & 1.19 & 1.37 & 1.60 & 1.93 & 2.29 & 1.06 & 1.18 & 1.42 & 1.81 & 2.29 & 0.033 \\
ResNet & 2.03 & \textbf{1.10} & 1.28 & 1.59 & 2.08 & 2.59 & 1.14 & 1.30 & 1.60 & 1.98 & 2.43 & 1.21 & 1.46 & 1.75 & 2.09 & 2.52 & 1.11 & 1.30 & 1.61 & 2.02 & 2.54 & 0.038\\
Prop. RDL-Net & 1.87 & \textbf{1.10} & \textbf{1.30} & \textbf{1.67} & \textbf{2.23} & \textbf{2.73} & \textbf{1.18} & \textbf{1.44} & \textbf{1.80} & \textbf{2.20} & \textbf{2.62} & \textbf{1.23} & \textbf{1.48} & \textbf{1.80} & \textbf{2.30} & \textbf{2.62} & \textbf{1.18} & \textbf{1.43} & \textbf{1.75} & \textbf{2.13} & \textbf{2.63} & 0.040 \\
\midrule
LSTM-IRM  & 30.7 & 1.07 & 1.20 & 1.46 & 1.88 & 2.31 & 1.08 & 1.17 & 1.40 & 1.71 & 2.13 & 1.09 & 1.24 & 1.46 & 1.71 & 2.00 & 1.06 & 1.18 & 1.40 & 1.72 & 2.12 & 0.030 \\
Xu2017 &  19.1 & \textbf{1.18} & \textbf{1.43} & 1.78 & 2.20 & 2.66 & 1.15 & 1.33 & 1.58 & 1.94 & 2.35 & 1.17 & 1.43 & 1.80 & 2.25 & 2.65 & 1.09 & 1.23 & 1.47 & 1.87 & 2.34 & 0.040 \\
Prop. RDL-Net & 3.91 & 1.13 & 1.36 & \textbf{1.79} & \textbf{2.46} & \textbf{2.98} & \textbf{1.19} & \textbf{1.42} & \textbf{1.83} & \textbf{2.27} & \textbf{2.74} & \textbf{1.26} & \textbf{1.53} & \textbf{1.86} & \textbf{2.31} & \textbf{2.78} & \textbf{1.19} & \textbf{1.46} & \textbf{1.83} & \textbf{2.26} & \textbf{2.74} & 0.045 \\
		\bottomrule          
	\end{tabular}
	\label{taba}
\end{table*}

\begin{table*}[h!]
	\centering
	\scriptsize
	\setlength{\tabcolsep}{3.6pt}
	\caption{Enhanced speech objective intelligibility scores (in $\%$) as given by the short-time objective intelligibility (STOI) metric \cite{short_obj}. The tested conditions include clean speech mixed with real-world \textbf{non-stationary} (\textit{voice babble} and \textit{street music}) and \textbf{coloured} (\textit{F16} and \textit{factory}) noise sources at multiple SNR levels. The highest STOI score attained at each condition and for each parameter size is shown in boldface. The standard error (SE) over all conditions for each network is provided in the last column.}
	
	\begin{tabular}{ll|lllll|lllll|lllll|lllll|l} 
		\toprule
		\multirow{3}{*}{\begin{tabular}[c]{@{}c@{}}\textbf{Network}\end{tabular}} &
		\multirow{3}{*}{\begin{tabular}[c]{@{}c@{}}\textbf{\# params.}\\$\pmb{\times 10^{6}}$\end{tabular}} &
		\multicolumn{20}{c|}{\bf SNR level (dB)} & \multirow{3}{*}{\begin{tabular}[c]{@{}c@{}}\textbf{SE}\end{tabular}} \\ 
		\cline{3-22}
		& & \multicolumn{5}{c|}{\bf Voice babble} & \multicolumn{5}{c|}{\bf Street music} & \multicolumn{5}{c|}{\bf F16} & \multicolumn{5}{c|}{\bf Factory} &  \\ 
		\cline{3-22}
		& & {\bf-5}  & {\bf0}   & {\bf5}   & {\bf10}  & {\bf15}        & {\bf-5}  & {\bf0}   & {\bf5}   & {\bf10}  & {\bf15}        & {\bf-5}  & {\bf0}   & {\bf5}   & {\bf10}  & {\bf15}  & {\bf-5}  & {\bf0}   & {\bf5}   & {\bf10}  & {\bf15} & \\ 
		\hline
Noisy speech & - & 60.2 & 72.4 & 83.0 & 90.7 & 95.5 & 59.0 & 70.9 & 81.9 & 90.3 & 95.6 & 60.4 & 71.8 & 82.4 & 90.5 & 95.7 & 57.8 & 69.9 & 80.9 & 89.2 & 94.5 & 0.010\\
\midrule
ResLSTM & 1.02 & 58.1 & 73.8 & 85.4 & 92.7 & 96.5 & 64.3 & 76.9 & 87.6 & 93.6 & 96.9 & 64.8 & 77.6 & 86.6 & 92.1 & 95.8 & 60.0 & 73.6 & 84.8 & 91.2 & 95.3 & 0.010 \\
DenseNet & 0.97 & 56.5 & 72.5 & 85.8 & 93.5 & 96.8 & 62.8 & 74.1 & 85.4 & 92.5 & 96.3 & 64.8 & 78.6 & 87.7 & 93.2 & 96.4 & 59.3 & 74.8 & 85.9 & 92.4 & 96.0 & 0.010\\
DenseRNet & 1.05 & 58.6 & 73.9 & 85.4 & 92.3 & 96.1 & 64.0 & 75.5 & 85.7 & 92.2 & 96.3 & 65.7 & 78.2 & 87.2 & 93.1 & 96.8 & 58.1 & 74.3 & 84.3 & 91.1 & 95.6 & 0.010 \\
ResNet & 1.03 & 59.8 & 75.4 & 87.4 & 94.0 & 97.0 & 65.4 & 77.2 & 88.0 & 94.0 & 97.2 & 68.0 & 80.3 & 88.7 & 94.1 & 97.1 & 62.5 & 76.9 & 86.7 & 92.9 & 96.4 & 0.009 \\
Prop. RDL-Net & 1.08 & \textbf{60.2} & \textbf{77.9} & \textbf{88.6} & \textbf{94.3} & \textbf{97.2} & \textbf{67.2} & \textbf{80.4} & \textbf{89.9} & \textbf{94.8} & \textbf{97.4} & \textbf{69.6} & \textbf{82.7} & \textbf{90.1} & \textbf{94.6} & \textbf{97.3} & \textbf{63.3} & \textbf{79.4} & \textbf{88.4} & \textbf{93.4} & \textbf{96.5} & 0.009 \\
\midrule
ResLSTM & 1.51 & 60.3 & 76.1 & 87.0 & 93.9 & 96.9 & 63.5 & 77.3 & 87.9 & 94.0 & 97.0 & 66.4 & 79.3 & 87.7 & 93.0 & 96.0 & 62.1 & 78.3 & 87.5 & 92.8 & 96.2 & 0.009 \\
DenseNet & 1.41 & 59.4 & 75.1 & 86.6 & 93.3 & 96.6 & 64.1 & 76.3 & 86.6 & 92.9 & 96.4 & 65.9 & 79.8 & 88.0 & 93.5 & 96.6 & 60.0 & 77.9 & 87.1 & 92.7 & 96.1 & 0.009 \\
DenseRNet & 1.37 &59.7 & 74.6 & 86.1 & 93.0 & 96.5 & 62.8 & 75.8 & 85.9 & 92.4 & 96.3 & 67.1 & 79.2 & 87.8 & 93.3 & 96.8 & 59.9 & 75.1 & 86.0 & 92.3 & 96.0 & 0.009 \\
ResNet & 1.53 & 60.9 & 76.5 & 87.9 & 94.0 & 97.1 & 66.0 & 77.9 & 88.2 & 93.8 & 97.0 & 67.9 & 80.9 & 89.3 & \textbf{94.3} & \textbf{97.3} & 63.2 & 78.3 & 87.8 & 93.1 & 96.5 & 0.009 \\
Prop. RDL-Net & 1.48 & \textbf{61.0} & \textbf{77.3} & \textbf{88.9} & \textbf{94.5} & \textbf{97.4} & \textbf{66.8} & \textbf{80.0} & \textbf{89.2} & \textbf{94.4} & \textbf{97.4} & \textbf{69.4} & \textbf{82.6} & \textbf{89.7} & \textbf{94.3} & 97.2 & \textbf{64.6} & \textbf{80.0} & \textbf{88.7} & \textbf{93.5} & \textbf{96.7} & 0.009 \\
\midrule
ResLSTM & 2.03 & 61.1 & 74.6 & 87.0 & 93.7 & 96.9 & 66.4 & 78.8 & 88.8 & 94.2 & 97.1 & 67.6 & 80.3 & 88.6 & 93.6 & 96.5 & 64.1 & 79.1 & 87.8 & 93.1 & 96.4 & 0.009 \\
DenseNet & 1.94 & 60.1 &75.3 & 86.9 & 93.9 & 96.9 & 64.8 & 77.1 & 86.9 & 93.2 & 96.8 & 66.7 & 79.9 & 88.3 & 93.3 & 96.5 & 60.4 & 77.2 & 87.4 & 92.8 & 96.3 & 0.009 \\
DenseRNet & 2.02 & 59.3 & 73.4 & 84.8 &92.0 & 95.8 & 64.0 & 74.8 & 84.4 & 91.2 & 95.5 & 66.5 & 77.8 & 86.4 & 92.4 & 96.1 & 58.6 & 73.4 & 84.1 & 91.2 & 95.4 & 0.009 \\
ResNet & 2.03 & \textbf{62.7} & 77.3 & 87.6 & 93.8 & 97.0 & 66.7 & 78.0 & 88.1 & 94.0 & 97.1 & 69.0 & 81.1 & 88.6 & 93.8 & 97.0 & 62.1 & 77.1 & 86.9 & 92.5 & 96.3 & 0.009 \\
Prop. RDL-Net & 1.87 & 61.5 & \textbf{77.8} & \textbf{89.0} & \textbf{94.7} & \textbf{97.4} & \textbf{68.5} & \textbf{81.2} & \textbf{90.1} & \textbf{94.8} & \textbf{97.4} & \textbf{69.3} & \textbf{82.5} & \textbf{90.4} & \textbf{94.9} & \textbf{97.4} & \textbf{64.8} & \textbf{80.6} & \textbf{88.6} & \textbf{93.5} & \textbf{96.6} & 0.009 \\
\midrule
LSTM-IRM & 30.7 & \textbf{64.2} & 78.5 & 88.0 & 93.5 & 96.5 & 66.1 & 77.4 & 86.6 & 92.6 & 96.0 & 67.3 & 79.1 & 87.3 & 92.5 & 95.8 & 62.3 & 76.7 & 86.6 & 92.5 & 95.9 & 0.009 \\
Xu2017 & 19.1 & 62.5 & 74.8 & 83.8 & 90.1 & 94.7 & 64.0 & 77.9 & 86.7 & 92.4 & 95.5 & 68.3 & 79.0 & 86.7 & 92.8 & 95.5 & 61.0 & 74.2 & 83.9 & 90.5 & 94.7 & 0.009 \\
Prop. RDL-Net & 3.91 & \textbf{64.2} & \textbf{80.82} & \textbf{89.0} & \textbf{94.7} & \textbf{97.4} & \textbf{71.6} & \textbf{82.9} & \textbf{90.7} & \textbf{95.0} & \textbf{97.5} & \textbf{72.6} & \textbf{83.9} & \textbf{91.0} & \textbf{95.4} & \textbf{97.8} & \textbf{67.1} & \textbf{81.7} & \textbf{89.5} & \textbf{93.9} & \textbf{96.8} & 0.008 \\
\bottomrule          
	\end{tabular}
	\label{tabb}
\end{table*}
\begin{table*}[ht]
	\centering
	\caption{Comparison to recent deep learning approaches to speech enhancement using the second test set. As in previous works, the objective scores are averaged over all tested conditions. \textbf{CSIG}, \textbf{CBAK}, and \textbf{COVL} are mean opinion score (MOS) predictors of the signal distortion, background-noise intrusiveness, and overall signal quality, respectively \cite{4389058}. \textbf{PESQ} is the perceptual evaluation of speech quality measure \cite{4389058}. \textbf{STOI} is the short-time objective intelligibility measure (in \%) \cite{short_obj}. The highest scores attained for each measure are indicated in boldface.}
\begin{tabular}{@{}llllll@{}}
\toprule
\textbf{Method}                            & \textbf{CSIG} & \textbf{CBAK} & \textbf{COVL} & \textbf{PESQ} & \textbf{STOI} \\ \midrule
Noisy speech & 3.35 & 2.44 & 2.63 & 1.97   & 92 (91.5)  \\
Wiener \cite{Wiener}           & 3.23          & 2.68          & 2.67          & 2.22   & -  \\
SEGAN \cite{segan}               & 3.48          & 2.94          & 2.80          & 2.16 & 93         \\
Wavenet \cite{wavenet}             & 3.62          & 3.23          & 2.98          & - & -            \\
MMSE-GAN \cite{MMSE-GAN}            & 3.80          & 3.12          & 3.14          & 2.53 & 93         \\
Deep Feature Loss \cite{deep_floss}   & 3.86          & 3.33 & 3.22          & -    & -         \\
Metric-GAN \cite{metric-GAN}            & 3.99          & 3.18          & 3.42          & 2.86 & -         \\ \midrule

Proposed RDL-Net 1.87M (Deep Xi - MMSE-LSA)                             & 4.29          & 3.32          & 3.62          & 2.93 & 93 (93.4) \\
Proposed RDL-Net 1.87M (Deep Xi - SRWF)                                 & 4.27 & 3.23          & 3.56 & 2.84 & 93 (93.5)\\ 
Proposed RDL-Net 3.91M (Deep Xi - MMSE-LSA)                             & \textbf{4.38}          & \textbf{3.43}          & \textbf{3.72}          & \textbf{3.02} & \textbf{94} (\textbf{93.8}) \\
Proposed RDL-Net 3.91M (Deep Xi - SRWF)                             & 4.36          & 3.35         & 3.67         & 2.94 & \textbf{94} (\textbf{93.8}) \\ \bottomrule
\end{tabular}
\label{table-stoa}
\end{table*}

\subsection{Training and validation error}
The training and validation error curves for the RDL-Net, ResNet, DenseNet, and DenseRNet at a parameter sizes of approximately 2 million are shown in Figures~\ref{fig:Curves} (a) and (b), respectively. The RDL-Net was able to converge to a lower training and validation error than the other networks. This suggests that the proposed RDL-Net allocated an efficient number of parameters for feature re-usage. Conversely, the DenseNet and DenseRNet struggled at a parameter size of 2 million, indicating that too many parameters were wasted on feature re-usage.  

\subsection{Parameter and computational efficiency}

The lowest validation error as a function of the number of parameters and computations for RDL-Nets, ResNets, DenseNets, and DenseRNets are shown in Figures\ref{fig:Curves} (c) and (d), respectively. RDL-Nets were able to achieve the same validation error as ResNets that employed significantly more parameters. For example, at a parameter size of 1 million, the RDL-Net attained the same lowest validation error as the ResNet with double the amount of parameters. A similar trend can be seen for the lowest validation error as a function of the number of FLOPs, (where FLOPs refers to the number of multiplication-addition operations during inference). For example, the RDL-Net that requires 2 million FLOPs achieved a lowest validation error similar to that of the ResNet that requires $4\times$ as many FLOPs.

\definecolor{darkgreen}{RGB}{0,128,0}
\begin{figure}[t]
\centering
\begin{tikzpicture}[scale=0.67]
\pgfplotstableread{error_results.txt}
    \datatable
    \begin{axis}[grid=major,ymin=140,ymax=148,xmin=2,xmax=100,xlabel= Epoch,
    ylabel= Training error,
    xlabel style={font=\normalsize}, xtick={20,40,60,80},
    title=(a), title style={yshift=-3mm,},
    ylabel style={font=\normalsize},yticklabel style={font=\normalsize},xticklabel style={font=\normalsize},
    width=6.2cm, height=5.5cm
    ]
    
	\addplot[line width=1pt, color=black, densely dotted] table[y = DenseRnet-tr] from \datatable ;
	
	\addplot[line width=1pt, color=darkgreen, densely dashdotted] table[y = DenseNet-tr] from \datatable ;
	
    \addplot[line width=1pt, color=blue, densely dashed] table[y = ResNet-tr] from \datatable ;
	
	\addplot[line width=1pt, color=red] table[y = RDL-tr] from \datatable ;
	\end{axis}
\end{tikzpicture}
\begin{tikzpicture}[scale=0.67]
\pgfplotstableread{error_results.txt}
    \datatable
    \begin{axis}[grid=major,ymin=140,ymax=148,xmin=2,xmax=100,xlabel= Epoch,
    ylabel= Validation error,legend pos=north east,
    xlabel style={font=\normalsize},  xtick={20,40,60,80},
    ylabel style={font=\normalsize},yticklabel style={font=\normalsize},xticklabel style={font=\normalsize},
    width=6.2cm, height=5.5cm, title=(b),  title style={yshift=-3mm,},
    ]
	
	\addplot[line width=1pt, color=black, densely dotted] table[y = DenseRnet-val] from \datatable ;
	\addlegendentry{DenseRNet}

	\addplot[line width=1pt, color=darkgreen, densely dashdotted] table[y = DenseNet-val] from \datatable ;
	\addlegendentry{DenseNet}
	
    \addplot[line width=1pt, color=blue, densely dashed] table[y = ResNet-val] from \datatable ;
	\addlegendentry{ResNet}
	
	\addplot[line width=1pt, color=red, mark repeat=15] table[y = RDL-val] from \datatable ;
	\addlegendentry{RDL-Net}
	
	\end{axis}
\end{tikzpicture}
\begin{tikzpicture}[scale=0.67]
	\begin{axis}[grid=major,xlabel= \# parameters ($\times 10^{6}$),
    ylabel= Lowest validation error,legend pos=north west,xlabel style={font=\normalsize}, ylabel style={font=\normalsize},yticklabel style={font=\normalsize},xticklabel style={font=\normalsize},
    width=6.2cm, height=5.5cm,
    title=(c), title style={yshift=-3mm,},
    ]
	\addplot[color=black,mark=diamond*, densely dotted, mark options={solid}, line width=1pt] coordinates {
	(0.6,  143.04)
	(1.05,  142.82)
	(1.37,  142.7)
	(2.02,  142.51)};
    \node [above,black] at (axis cs:  1.5,  142.7) {DenseRNet};
    
	\addplot[color=darkgreen,mark=*, densely dashdotted, mark options={solid}, line width=1pt] coordinates {
	(0.57,  143.25)
	(0.97,  142.8)
	(1.48,  142.51)
	(2.1,  142.21)};
    \node [above,darkgreen] at (axis cs:  1.5,  142) {DenseNet};
	
	\addplot[color=blue,mark=square*,densely dashed, mark options={solid}, line width=1pt] coordinates {
	(0.53,  142.19)
	(1.02,  141.82)
	(1.51,  141.42)
	(2.03,  141.11)};
    \node [right,blue] at (axis cs:  0.53,  142.29) {ResNet};   
    
	\addplot[color=red,mark=triangle*,mark options={solid}, line width=1pt] coordinates {
    (0.53,  142.14)
	(1.08,  141.03)
	(1.48,  140.7)
	(1.92,  140.5)};
    \node [right,red] at (axis cs:  0.53,  140.8) {RDL-Net}; 	
    \draw[dashed,->,very thick] (axis cs:2,  141.11) -- (axis cs:1.03,  141.11);
    
    \node [right,black] at (axis cs:  0.935,  141.26) {\textbf{$\approx $2$\pmb{\times}$ fewer params}}; 
    
	\end{axis}
\end{tikzpicture}
\begin{tikzpicture}[scale=0.67]
	\begin{axis}[grid=major,xlabel= \# FLOPs ($\times 10^{6}$),
    ylabel= Lowest validation error,legend pos=north west,xlabel style={font=\normalsize}, ylabel style={font=\normalsize},yticklabel style={font=\normalsize},xticklabel style={font=\normalsize},
    width=6.2cm, height=5.5cm,     title=(d), title style={yshift=-3mm,},
    ]

	\addplot[color=black,mark=diamond*, densely dotted, mark options={solid}, line width=1pt] coordinates {
	(2.9,  143.04)
	(4.2,  142.82)
	(5.48,  142.7)
	(8.06,  142.51)};
    \node [above,black] at (axis cs:  7.5,  142.6) {DenseRNet};

	\addplot[color=darkgreen,mark=*, densely dashdotted,mark options={solid}, line width=1pt] coordinates {
	(2.26,  143.25)
	(3.87,  142.8)
	(5.9,  142.52)
	(8.34,  142.21)};
    \node [below,darkgreen] at (axis cs:  7.5,  142.23) {DenseNet};
	
	\addplot[color=blue,mark=square*, densely dashed, mark options={solid}, line width=1pt] coordinates {
	(2.48,  142.19)
	(4.56,  141.82)
	(6.59,  141.42)
	(8.71,  141.11)};
    \draw[dashed,->,very thick] (axis cs:8.71,  141.11) -- (axis cs:2.16,  141.11);
    \node [right,blue] at (axis cs:  2.5,  142.25) {ResNet};   
    
    \node [right,black] at (axis cs:  2.2,  141.24) {\textbf{$\approx $4$\pmb{\times}$ fewer FLOPs}}; 	
    
	\addplot[color=red,mark=triangle*, mark options={solid}, line width=1pt] coordinates {
    (1.23,  142.14)
	(2.16,  141.03)
	(2.9,  140.7)
	(3.7,  140.5)};
    \node [right,red] at (axis cs:  3.7,  140.54) {RDL-Net}; 	
	\end{axis}

\end{tikzpicture}
\caption{Training plots for RDL-Nets, ResNets, DenseNets and DenseRNets: \textbf{(a)} training error, \textbf{(b)} validation error, and lowest validation error as a function of the number of \textbf{(c)} parameters and \textbf{(d)} FLOPs.}
	\label{fig:Curves}
\end{figure}
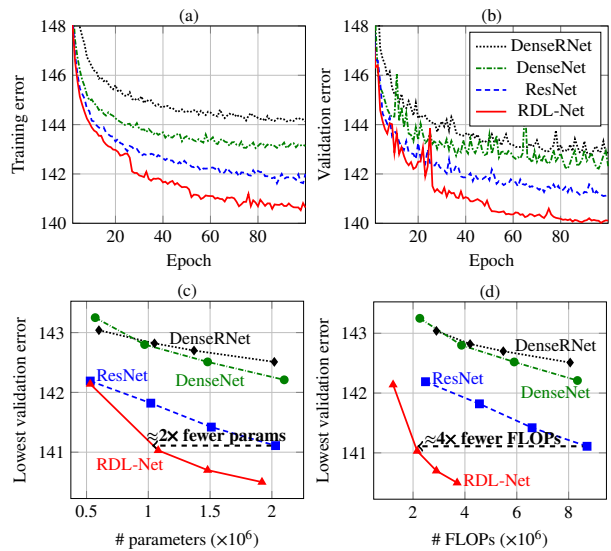

\subsection{Speech enhancement performance}
The enhanced speech objective quality scores attained by each of the networks in the Deep Xi framework are presented in Table \ref{taba}. Each network estimated the \textit{a priori} SNR for the MMSE-LSA estimator. It can be seen that RDL-Nets were able to achieve the highest objective quality scores for most of the tested conditions. The performance capability of RDL-Nets was demonstrated at a parameter size of 2 million for \textit{street music} at 10 dB, where the RDL-Net achieved a MOS-LQO improvement of 0.22 over the ResNet. Table~\ref{tabb} shows the objective intelligibility scores obtained by each of the networks. It can be seen that RDL-Nets were able to achieve the highest objective intelligibility scores for most of the tested conditions. RDL-Nets demonstrated its performance at a parameter size of 2 million for~\textit{factory} noise at 0 dB, attaining an STOI improvement of $3.5\%$ when compared to the equivalent ResNet. RDL-Nets in the Deep Xi framework were also able to produce enhanced speech with higher objective quality and intelligibility scores than two other widely known deep learning speech enhancement frameworks (LSTM-IRM and Xu2017) \cite{xu2017multi,chen2017long}.

\begin{figure} [h!]
	\begin{center}
		\includegraphics[scale=0.76]{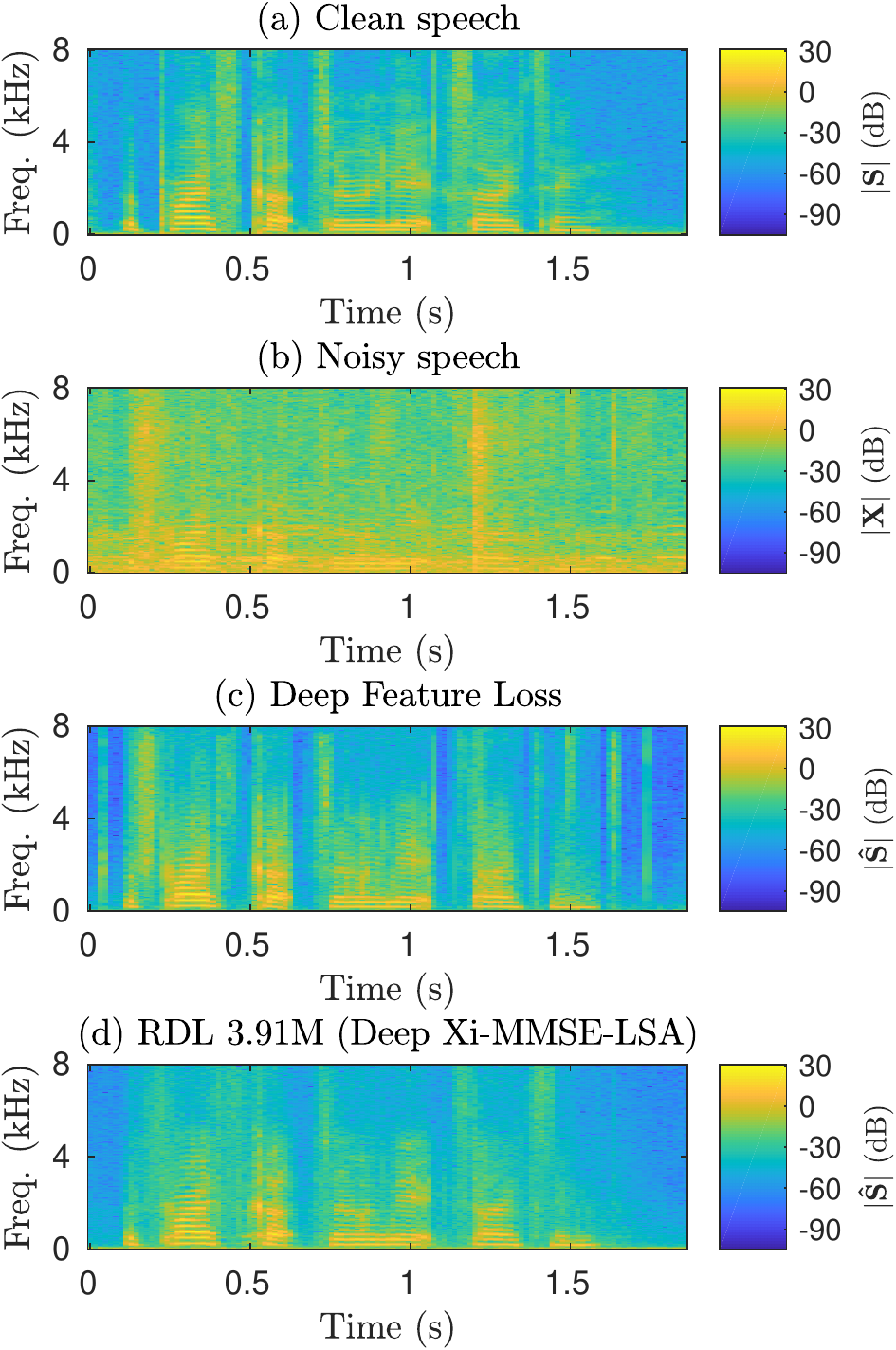}
		\caption{(a) Clean speech magnitude spectrogram ($|\textbf{S}|$) of female $p257$ uttering sentence $70$, ``The price cuts are really exciting''. (b) \textit{Crowd noise} mixed with (a) at an SNR level of 2.5 dB ($|\textbf{X}|$). Enhanced speech ($|\hat{\textbf{S}}|$) produced by (c) Deep Feature Loss and (d) RDL-Net 3.91M (Deep Xi-MMSE-LSA).}
		\label{fig-spectogram} 
	\end{center}
\end{figure}

We also compare RDL-Nets to recent deep learning approaches to speech enhancement. Here, RDL-Nets were used to estimate the \textit{a priori} SNR for the SRWF approach and the MMSE-LSA estimator. As shown in Table \ref{table-stoa}, RDL-Nets were able to attain the highest CSIG, CBAK, COVL, PESQ and STOI scores. The RDL-Net demonstrated an improvement of 0.39, 0.25, 0.3, and 0.16 over Metric-GAN for CSIG, CBAK, COVL, and PESQ, respectively. The RDL-Net also demonstrated an improvement of $1\%$ over MMSE-GAN for STOI. The enhanced speech produced by RDL-Net 3.91M is illustrated in Figure \ref{fig-spectogram} (d). It can be seen that the RDL-Net demonstrated superior noise suppression with little formant distortion. As illustrated in Figure \ref{fig-spectogram} (c), Deep Feature Loss over- and under-estimated multiple spectral components.\footnote{Enhanced speech recordings and additional results are available at: \url{https://github.com/nick-nikzad/RDL-SE}.}

\section{Conclusion} \label{secd}
In this paper, we propose a novel convolutional neural network (CNN) for speech enhancement, called a residual-dense lattice (RDL) network. Unlike other CNNs that use both residual and dense aggregations, RDL-Nets take advantage of both aggregation types without over-allocating parameters for feature re-usage. This enables RDL-Nets to produce a higher speech enhancement performance than other networks, such as ResLSTM networks, ResNets, DenseNets, and DenseRNets. We also show that RDL-Nets are able to outperform many state-of-the-art deep learning approaches to speech enhancement. In future work, the RDL-Net topology will be investigated for speech separation, speech recognition, computer vision, and image denoising.

\bibliographystyle{aaai}
\bibliography{ms}
\end{document}